# Properties of Digital n-Dimensional Spheres and Manifolds. Separation of Digital Manifolds

Alexander V. Evako
Npk Novotek , Moscow, Russia
e-mail: evakoa@mail.ru.

**Abstract**

In the present paper, we study basic properties of digital n-dimensional manifolds and digital simply connected spaces. An important property of a digital n-manifold is that M is a digital n-sphere if and only if for any point v of M, M-v is a digital n-disk. It is proved that a digital (n-1)-sphere S contained a digital n-sphere M is a separating space of M. We show that a digital n-manifold can be converted to the compressed form by sequential contractions of simple pairs of adjacent points. We study structural features of digital simply connected spaces. In particular, a digital (n-1)-sphere S in a digital simply connected n-manifold M is a separating space of M.

**1. Introduction**

Topological properties of two-and three-dimensional image arrays play an important role in image processing operations. A consistent theory for studying the topology of digital images in n dimensions can be used in a range of applications, including pattern analysis, medical imaging, computer graphics, detection of dynamically moving surfaces, and representation of microscopic cross-sections.

In recent years, there has been a considerable amount of works devoted to investigating digital spaces. X. Daragon et al. [4-5] studied partially ordered sets in connection with the notion of n-surfaces. In particular, it was proved that (in the framework of simplicial complexes) any n-surface is an n-pseudomanifold, and that any n-dimensional combinatorial manifold is an n-surface. In papers [7-8], a digital n-surface was defined and basic properties of n-surfaces were studied. Paper [7] analyzes a local structure of the digital space $Z^n$. It is shown that $Z^n$ is an n-surface for all n>0. In paper [8], it is shown that if A and B are n-surfaces and A⊆B, then A=B. .In paper [20], M. Smyth et al. defined dimension at a vertex of a graph as basic dimension, and the dimension of a graph as the sup over its vertices. They proved that dimension of a strong product G × H is dim ( G ) + dim ( H ) (for non-empty graphs G and H). An interesting method using cubical images with direct adjacency for determining such topological invariants as genus and the Betti numbers was designed and studied by L. Chen et al. [3]. E. Melin [14] studies the join operator, which combines two digital spaces into a new space. Under the natural assumption of local finiteness, he shows that spaces can be uniquely decomposed as a join of indecomposable spaces. Digital simple closed curves were studied in [1, 13]. It was shown that a digital simple closed curve of more than four points is not contractible.

In the present paper, we focus on digital spaces with properties, which closely resemble properties of their continuous counterparts. In sections 3, digital n-spheres and n-manifolds are investigated. We show that a digital n-manifold M is a digital n-sphere if for any point v, the space M-v is a digital n-disk. It is proven that a digital (n-1)-sphere S contained in a digital n-sphere M is a separating space of M. We study compressed digital n-manifolds and show that if a compressed n-manifold M contains more them 2n+2 points then M is not a digital n-sphere.

Section 4 introduces the notion of a digital simple closed curve (which is different from the notions of simple closed curves given in [1] and [13]) and defines a simply connected digital space, and a locally simply connected digital space. We prove that if n-manifolds M and N are homotopy equivalent and M is locally simply connected then so is N. It is shown that if a digital n-manifold M is locally simply connected and S is a digital (n-1)-sphere lying in M, then S is a separating space of M. We prove that a locally simply connected digital 2-manifold is a digital 2-sphere. The main result obtained in this section says that a digital locally simply connected 3-manifold is a digital 3-sphere.

## 2. Digital spaces, contractible graphs and contractible transformations

A digital space G is a simple undirected graph $G=(V,W)$, where $V=\{v_1,v_2,...v_n,...\}$ is a finite or countable set of points, and $W = \{(v_p v_q),....\} \subseteq V \times V$ is a set of edges. Such notions as the connectedness, the adjacency, the dimensionality and the distance on a graph G are completely defined by sets V and W (see, e.g., [4-11, 14, 20]).

We use the notations $v_p \in G$ and $(v_p v_q) \in G$ if $v_p \in V$ and $(v_p v_q) \in W$ respectively if no confusion can result. $|G|$ denotes the number of points in G.

Since in this paper we use only subgraphs induced by a set of points, we use the word subgraph for an induced subgraph. We write $H \subseteq G$. Let G be a graph and $H \subseteq G$. G-H will denote a subgraph of G obtained from G by deleting all points belonging to H. For two graphs $G=(X,U)$ and $H=(Y,W)$ with disjoint point sets X and Y, their join $G \oplus H$ is the graph that contains G, H and edges joining every point in G with every point in H. The subgraph $O(v) \subseteq G$ containing all points adjacent to v (without v) is called the rim or the

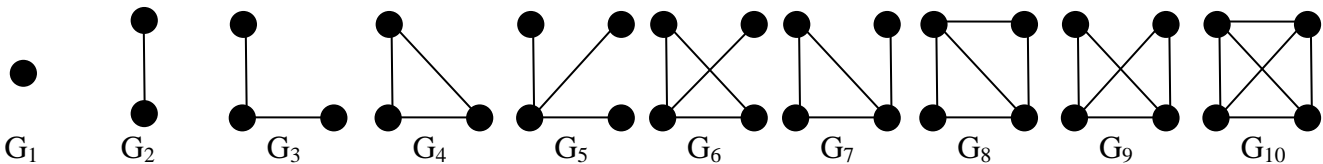

G₁  G₂  G₃  G₄  G₅  G₆  G₇  G₈  G₉  G₁₀

Figure 1. Contractible graphs with the number of points n<5.

neighborhood of point v in G, the subgraph $U(v)=v \oplus O(v)$ is called the ball of v. Graphs can be transformed from one into another in a variety of ways. Contractible transformations of graphs seem to play the same role in this approach as a homotopy in algebraic topology [10-11].

A graph G is called contractible (fig. 1), if it can be converted to the trivial graph by sequential deleting simple points. A point v of a graph G is said to be simple if its rim $O(v)$ is a contractible graph. An edge (vu) of a graph G is said to be simple if the joint rim $O(vu)=O(v) \cap O(u)$ is a contractible graph. In [10], it was shown that if (vu) is a simple edge of a contractible graph G, then G-(vu) is a contractible graph. Thus, a contractible graph can be converted to a point by sequential deleting simple points and edges. In fig.1, $G_{10}$ can be converted to $G_9$ or $G_8$ by deleting a simple edge. $G_9$ can be converted to $G_7$ or $G_6$ by deleting a simple edge. $G_6$ can be converted to $G_5$ by deleting a simple edge. $G_7$ can be converted to $G_4$ by deleting a simple point. $G_5$ can be converted to $G_3$ by deleting a simple point. $G_3$ can be converted to $G_2$ by deleting a simple point. $G_2$ can be converted to $G_1$ by deleting a simple point.

Deletions and attachments of simple points and edges are called contractible transformations. Graphs G and H are called homotopy equivalent or homotopic if one of them can be converted to the other one by a sequence of contractible transformations.

Homotopy is an equivalence relation among graphs. Contractible transformations retain the Euler characteristic and homology groups of a graph [11].

Properties of graphs that we will need in this paper were studied in [8-11].

**Proposition 2.1**
- Let G be a graph and v be a point ($v \notin G$). Then the cone $v \oplus G$ is a contractible graph.
- Let G be a contractible graph and S(a,b) be a disconnected graph with just two points a and b. Then $S(a,b) \oplus G$ is a contractible graph.
- Let G be a contractible graph with the cardinality $|G|>1$. Then it has at least two simple points.
- Let H be a contractible subgraph of a contractible graph G. Then G can be transformed into H by sequential deleting simple points.
- Let graphs G and H be homotopy equivalent. G is connected if and only if H is connected. Any contractible graph is connected.

In graph theory, the contraction of points x and y in a graph G is the replacement of x and y with a point z such that z is adjacent to the points to which points x and y were adjacent. In paper [10], the contraction of simple pairs of points  was used for classification of digital n-manifolds.

**Definition 2.1.**
- Let G be a graph and x and y be adjacent points of G. We say that {x,y} is a simple pair  if any point v belonging to U(x)-U(y) is not adjacent to any point u belonging to U(y)-U(x).
- Let G be a graph  and  {x,y} be a simple pair of G. The replacement of x and y with a point z such that  O(z)=U(x)∪U(y)-{x,y} is called the simple contraction of points x and y or C-transformation. CG=(G∪z)-{x,y} is the graph that results from contracting points x and y.
- Let G be a graph  and  z be a point of G. The replacement of z with adjacent points  x and y in such a way that U(x)∪U(y)-{x,y}=O(z), and  any point v belonging to U(x)-U(y) is not adjacent to any point u belonging to U(y)-U(x) is called the simple splitting of z or R-transformation. RG=(G∪{x,y})-z is the graph that results from simple splitting point z.

Simple C- and R-transformations are invertible.  For a given C-transformation, the  inverse of C is a simple splitting R=C$^{-1}$. In fig. 2(a), {x,y} is a simple pair of a digital 1-sphere M.  In fig. 2(b-c), (M∪z)-{x,y}is a digital 1-sphere obtained by {x,y} contraction. A pair {x,y} is simple in a digital 2-sphere (see fig. 2 (d-f)). A pair {a,b} depicted  in fig. 2(g) is  not a simple pair.

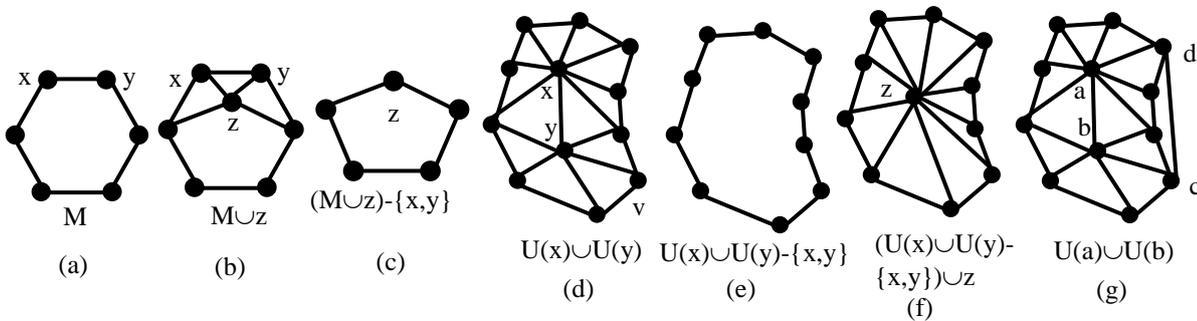

Figure  2.  (a) M is a digital 1-sphere with six points. (b) {x,y} is a topological pair of points. (c) N=(M∪z)-{x,y}is a digital 1-sphere with five points. (d)-f)  U(x)∪U(y) is a contractible space. U(x)∪U(y)-{x,y} is a digital 1-sphere. {x,y} is a topological pair of points.  (g) {a,b} is not a topological pair of points. U(a)∪U(b)-{a,b} is not a digital 1-sphere.

**Proposition 2.2.**
Let {x,y} be  a simple  pair lying in a graph G.   Then the graph H=(G-{x,y})∪z obtained by  the contraction of {x,y} is homotopy equivalent to G.
**Proof.**
First, show that the graph B=U(x)∪U(y) is contractible. Pick a point v∈U(x)-U(y). Since v is not adjacent to any point u belonging to U(y)-U(x) then the rim O$_B$(v) of v is the cone x⊕(O(xv), i.e., a contractible graph. Therefore, v is a simple point of B, and can be deleted from B. For the same reason, all points belonging to U(x)-U(y) can be deleted from B by sequential deleting simple point. The obtained graph U(y)=y⊕(O(y)  is homotopy equivalent to B. Since U(y) is a contractible graph according to proposition 2.1, then B=U(x)∪U(y) is a contractible graph.
Glue a simple point z to G in such a way that O(z)=U(x)∪U(y). In the obtained graph P=G∪z, the rim of x is the cone O$_P$(x)=z⊕O(x).  Therefore, point x is simple in P and can be deleted from P. In the obtained graph Q=P-{x}, the rim of y is the cone O$_Q$(y)=z⊕(O(y)-{x}). Therefore, y is simple in Q and can be deleted from Q. The obtained graph Q-{y}=H=(G-{x,y})∪z is homotopy equivalent to G. The  proof  is  complete.☐

**Proposition 2.3.**
Let {x,y} be  a simple  pair of a graph G.  Let H be a subgraph of G containing {x,y}.  Then {x,y} is a simple pair of H.
**Proof.**
Evidently, U$_H$(x)=U(x)∩H, U$_H$(y)=U(y)∩H. Since any point v belonging to U(x)-U(y) is not adjacent to any point u belonging to U(y)-U(x), then any point v belonging to (U(x)-U(y))∩H  is not adjacent to any point u belonging to (U(y)-U(x))∩H. Therefore, {x,y} is a simple pair of H.  The  proof  is  complete.  ☐

Further on, if we consider a graph together with the natural topology on it, we will use the phrase 'digital space". We say "space" to abbreviate "digital space", if no confusion can result.

### 3. Digital n-dimensional spheres

There is an abundant literature devoted to the study of different approaches to digital lines surfaces and spaces used by researchers, just mention some of them [1, 3-6, 20]. In paper [7], digital n-surfaces were introduced and studied. A digital 0-dimensional surface is a disconnected graph $S^0(a,b)$ with just two points a and b. A connected digital space M is called a digital n-dimensional surface, n>1, if the rim O(v) of any point v is a digital (n-1) surface.

**Definition 3.1.**
The join $S^n_{min}=S^0_1 \oplus S^0_2 \oplus \ldots S^0_{n+1}$ of (n+1) copies of the zero-dimensional surface $S^0$ is called a minimal n-sphere (see fig.3).

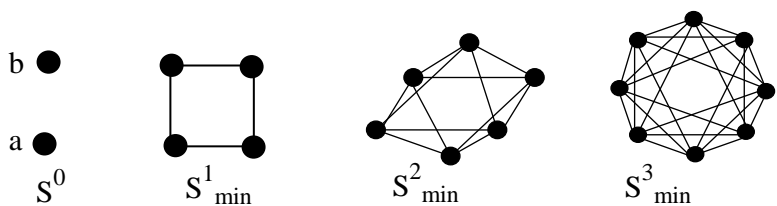

Figure 3. Minimal 1-, 2- and 3-dimensional spheres.

Evidently, $S^n_{min}$ is an n-surface containing 2n+2 points. Fig. 3 illustrates definition 3.1, and shows minimal spheres. A digital n-manifold is a special case of a digital n-surface defined and investigated in [7]. To define digital n-spheres, n>0, we will use a recursive definition. Suppose that we have defined digital k-spheres for dimensions $0 \leq k \leq n-1$.

**Definition 3.2.**
A connected digital space M is called a digital n-sphere, n>0, if for any point v of M, the rim O(v) is a digital (n-1)-sphere, and M can be converted to a minimal n-sphere $S^n_{min}$ by sequential contractions of simple pairs.

Note that a digital n-sphere is homotopy equivalent to a minimal n-sphere according to proposition 2.2. Digital 1- and 2-spheres are depicted in fig. 4 and 5. In fig. 2(a), {x,y} is a simple pair of a digital 1-sphere

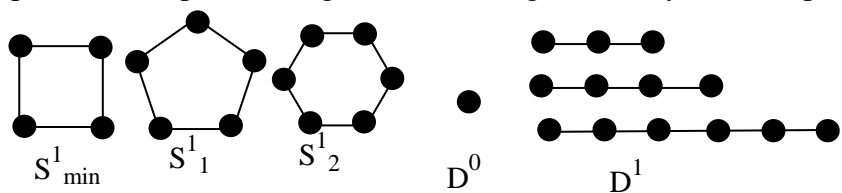

Figure 4. Zero- and one-dimensional spheres $S^0$ and $S^1$ and zero- and one-dimensional disks $D^0$ and $D^1$.

M. In fig. 2(b-c), $(M \cup z)$-{x,y} is a digital 1-sphere obtained by {x,y} contraction. A pair {x,y} is simple in a digital 2-sphere (see fig. 2 (d-f)). A pair {a,b} depicted in fig. 2(g) is not a simple pair. Notice that according to definition 2.1, a pair {x,y} of adjacent points x and y lying on a digital n-sphere M is a simple if any point v belonging to U(x)-U(y) is not adjacent to any point u belonging to U(y)-U(x), i.e., U(y)-U(x)-{x,y}=(O(x)-y)∪(O(y)-x) where O(x) and O(y) are digital (n-1)-spheres. For a simple {x,y} contraction, there exists the inverse transformation, which is the replacement of z with two adjacent points x and y such that O(x) and O(y) are digital (n-1)-spheres, and O(x)∪O(y)-{x,y}=O(z). A minimal n-sphere has no simple pairs, and according to definition 2.1, $S^n_{min}$ can be converted to M by a sequence of splittings, i.e., if $S^n_{min}=C_p \ldots C_1 M$ then $M=R_1 \ldots R_p S^n_{min}$, $R_i=C_i^{-1}$, i=1,…p. The following proposition shows the structure of a simple pair of points lying in a digital n-sphere.

## Proposition 3.1.
Let M be a digital n-sphere, n>0, {x,y} be a simple pair lying in M, and N=(M∪z)-{x,y} be the space obtained by the contraction of {x,y}. Then U(x)∪U(y)-{x,y} is a digital (n-1)-sphere, and N=(M∪z)-{x,y} is a digital n-sphere.

To make the reading easier, we have presented all proofs, except that of proposition 3.4 below, in Appendix 1. Digital 2-spheres are shown in fig. 5. $S^2_2$ contains a simple pair {x,y}. Contracting {x,y} converts $S^2_2$ to $S^2_{min}$.

## Definition 3.3.
Let M be a digital n-sphere, n>0, and v be a point belonging to M. Then the space D=∂D∪IntD=M-v is called a digital n-disk, ∂D=O(v) is called the boundary of D, and IntD is called the interior of D (see fig. 4, 5).

The following corollary is a consequence of definition 3.3.

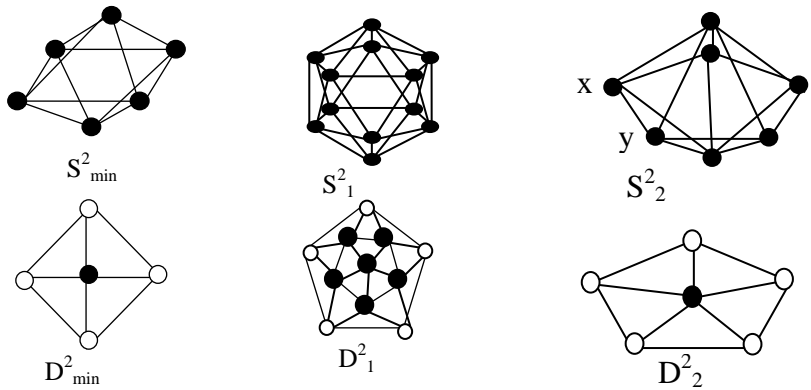

Figure 5. Digital 2-spheres and 2-disks.

## Remark 3.1.
Let D=∂D∪IntD be a digital n-disk D. Then ∂D is a digital (n-1)-sphere, the rim O(x) is a digital (n-1)-disk if x∈∂D (fig. 4, 5), and the rim O(x) is a digital (n-1)-sphere if a point x∈IntD.

## Proposition 3.2.
Let M be a digital n-sphere, n>0, and v be a point belonging to M. A digital n-disk D=M-v is a contractible space.

## Proposition 3.3.
Let M be a digital n-sphere, n>0, v be a point belonging to M, and M-v=D=∂D∪IntD be a digital n-disk. If |IntD|>1, then IntD contains a simple pair.

## Remark 3.2.
Notice that if {x,y} is a simple pair lying in a digital n-sphere M, then the union U(x)∪U(y) is a digital n-disk D=∂D∪IntD with the boundary ∂D=U(x)∪U(y)-{x,y} and IntD=[x,y], as it follows from definition 3.3 and proposition 3.4. Let M be a digital n-sphere, and D=∂D∪IntD be a digital n-disk lying in M. The replacement of IntD with a point z such that O(z)=∂D is called the contraction of D in M. Obviously, if M is a digital n-sphere. and D=∂D∪IntD is a digital n-disk lying in M, then the space N=(M∪z)-IntD obtained by the contraction of IntD is a digital n-sphere.

**Proposition 3.4.** Let M be an n-sphere and G be a contractible space contained in M. Then the space M-G is a contractible space.
**Proof**. The proof is by induction on the dimension n. For n=1, the proposition is verified directly. Assume that the proposition is valid whenever n<k+1. Let n=k+1. Since G is contractible, there is a point x belonging to G and simple in G, i.e., O(x)∩G is contractible according to proposition 2.1. Since O(x) is an

(n-1)-sphere, then by the induction hypothesis, O(x)-G=O(x)∩(M-G) is also contractible. Hence, x is simple in M-G. Therefore, $G_1$= G-x is a contractible space and M-$G_1$=(M-G)∪x is homotopy equivalent to M-G. Acting in the same way we finally convert the space G to a point v and the space M-G to the space M-v. Spaces M-v and M-G are homotopy equivalent by construction. Since M-v is contractible then M-G is a contractible space. This completes the proof. □

**Corollary 3.1.**
As it follows from proposition 3.4, if D=∂D∪IntD is a digital n-disk, then IntD is a contractible space. Indeed, D=M-v, where M is a digital n-sphere and v is a point of M. The ball U(v) of v is a contractible space. Therefore, M-U(v)=IntD is a contractible space.

**Proposition 3.5.**
(1) Let M be a digital n-sphere, n>0, G be a digital (n-k)-sphere lying in M, {x,y} be a simple pair lying in M, and N=FM=(M∪z)-{x,y} be the space obtained from M by the contraction of {x,y}. Then $G_1$=FG is a digital (n-k)-sphere lying in N.
(2) Let D be a digital n-disk, and {x,y} be a simple pair of adjacent points lying in D. Then E=FD=(D∪z)-{x,y} is a digital n-disk.

## 4. Structural properties of digital n-spheres. Separation of spaces.

**Definition 4.1.**
Let A and B be subspaces of a connected digital space M. A and B are called *separated* in M if any point of A is non-adjacent to any point of B. We will say that the union M=A∪C∪B is a *separation* of M by the space C and C is a *separating space* for M (see fig. 6(b-c)).

In fig. 6 (a), a digital 0-sphere $S^0$={x,y} is a separating space of a digital 1-sphere $S^1$. A digital 1-sphere C (see fig. 6 (d)-(e)) is a separating space of digital 2-spheres S(7) and S(9).

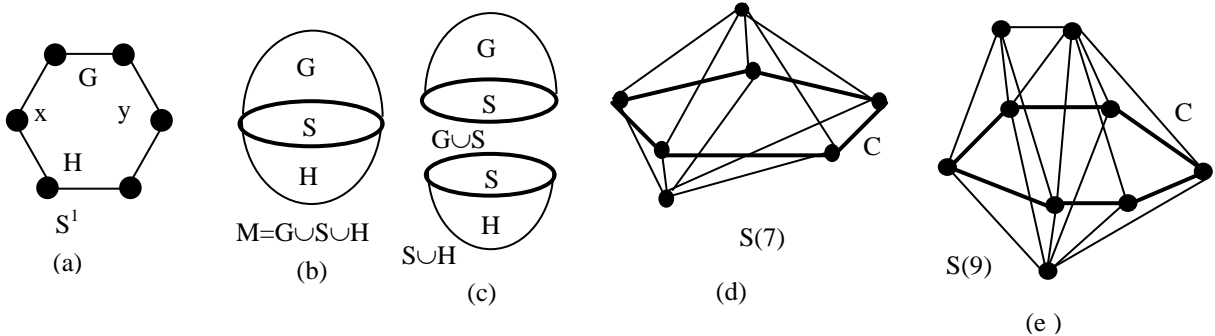

Figure 6. (a) A digital 0-sphere $S^0$={x,y} is a separating space in a 1-sphere $S^1$. (b) The separation of M by S. (c) G∪S and S∪H are digital n-disks. (d)- (e) A digital 1-sphere C is a separating space in 2-spheres S(7) and S(9).

**Proposition 4.1.**
Let M be a digital n-sphere, and S be a digital (n-1)-sphere lying in M. Then S is a separating space of M=A∪S∪B, and A∪S and S∪B are digital n-disks.
**Proof.** The proof is by induction on the number of points |M|=k of M. For k=2n+2, M is a minimal digital n-sphere M=$S^0_1$⊕$S^0_2$⊕…$S^0_{n+1}$=$S^0$(v,u)⊕$S^0_2$⊕…$S^0_{n+1}$= $S^0$(v,u)⊕$S^{n-1}_{min}$ =v∪ $S^{n-1}_{min}$ ∪u.
Assume that the proposition is valid whenever k<s. Let k=s. With no loss of generality, suppose that a simple pair {x,y}⊆A and {x,y}∩S=∅. Then N=(M∪{z})-{x,y} is a digital n-sphere, S⊆N, and |N|=s-1. Therefore, N=G∪S∪H, where S is a separating space, and G∪S and S∪H are digital n-disks by the induction hypothesis. Suppose that a point z belongs to G. Since G∪S=((A∪S)∪{z})={x,y}, then A∪S is homotopy equivalent to G∪S, i.e., a digital n-disk according to propositions 3.3. For the same reason, B∪S is a digital n-disk. The proof is complete. □

**Proposition 4.2.**
  (a) Let M be a digital 2-sphere, and |M|=7. Then M can be represented as the separation M=A∪S∪B, where |A|=2, |B|=1, and S is a digital minimal 1-sphere.
  (b) Let M be a digital 2-sphere, and |M|>7. Then M can be represented as the separation M=A∪S∪B, where |A|>1, |B|>1, and S is a digital 1-sphere.

**Proof.**
Note first that for a digital 1-sphere C with the number of points |C|>5, C=A∪$S^0$∪B is a separation of C by a 0-sphere $S^0$ with |A|>1 and |B|>1 (see fig. 6(a)).
( a) It is easy to check directly that if |M|=7, then M=$S^0_1$(x,y)⊕$S^1$(5), where $S^1$(5) is a digital 1-sphere {a,b,c,d,e} consisting of five points (see fig. 7(a)). Therefore, M= A∪C∪B, where |A|={e}, B={b,c}, and C={a,x,d,y}, is a digital 1-sphere.

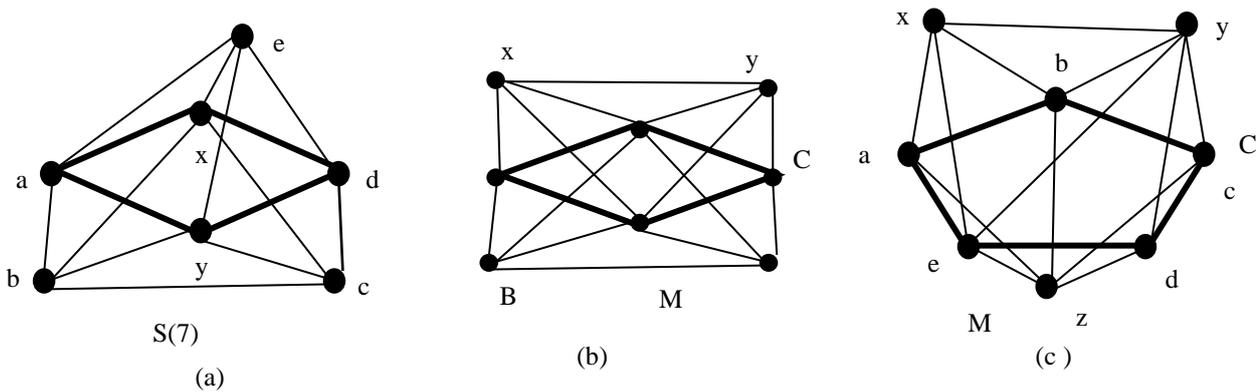

Figure 7. (a-c) Digital 2-spheres.

( b) Let |M|>7, and M contain a simple pair {x,y}, i.e., C=U(x)U(y)-{x,y} is a digital 1-sphere.
Suppose that B=M- U(x)U(y) contains several points, |B|>1. Then M= A∪C∪B, where |B|>1, |A|={x,y}, is a separation of M by C (fig. 7(b)).
Suppose that B=M- U(x)U(y) contains the only point z (see fig. 7(c )). Then |C|>4. For simplicity, suppose that C={a,b,c,d,e}. Then $C_1$={y,e,z,b} is a digital 1-sphere, B={a,x}, and M= A∪$C_1$∪B is a separation of M by $C_1$. The proof is complete. □

## 5. Digital n-dimensional manifolds

**Definition 5.1.**
- A connected space M is called a digital n-dimensional manifold, n>1, if the rim O(v) of any point v is a digital (n-1) sphere.
- Let M be an n-manifold and a point v belong to M. Then the space M-v=N=∂N∪IntN is called an n-manifold with the spherical boundary ∂N=O(v) and the interior IntN=N-∂N=M-U(v)..

Further on in this paper, we will study only n-manifolds with the spherical boundary.. Obviously, a digital n-sphere is a digital n-manifold. A digital 2-dimensional torus T and a digital 2-dimensional projective plane P are depicted in fig. 8. T-{7} and P-{a} are 2-manifolds with spherical boundaries O({7}) and O({a}) respectively.

Let N be an n-manifold with the (spherical) boundary ∂N and the interior IntN. Then ∂N is a digital (n-1)-sphere, the rim O(x) is a digital (n-1)-disk if x∈∂N, and the rim O(x) is a digital (n-1)-sphere if a point x∈IntN (T-{7} and P-{a} in fig. 8).

**Proposition 5.1.**
Let M be an n-manifold, G and H be contractible subspaces of M, and v be a point in M. Then subspaces M-G, M-H and M-v are all homotopy equivalent to each other.
**Proof**. Notice that repeating word for word the proof of proposition 3.4, we show that M-G is homotopy equivalent to M-v, where v is a point belonging to G. Similarly, M-H is homotopy equivalent to M-u, where u is a point belonging to H. Consider a path P(v,u) connecting points v and u. Since P is a contractible space, then for the same reason as above, M-P, M-v and M-u are homotopy equivalent. Hence, M-G, M-H, M-v and M-u are all homotopy equivalent. □

Let us emphasize that the following assertion summarizes previous results and shows a major difference between n-manifolds-spheres and n-manifolds-non-spheres.

**Corollary 5.1.**
Let M be a digital n-manifold. M is a digital n-sphere if and only if for any contractible space G contained in M, the space M-G is contractible.

A digital 2-torus T is shown in fig. 8. T-{7} (fig. 8) is homotopy equivalent to the space E (fig. 8) which is not contractible. P (fig. 8) is a digital 2-dimensional projective plane. It is easy to check directly P-{a} (fig. 8) is homotopy equivalent to the space C (fig. 8), which is a digital 1-sphere.
In a common sense, a digital n-dimensional sphere $S^n$ is the simplest n-manifold since it contains the smallest number of points compared to any other n-manifold [10].

As it is for a digital n-sphere, if a pair {x,y} of adjacent points x and y lying in a digital n-manifold M is simple then the contraction of {x,y} does not change the topology of the manifold.

The following assertion is a direct consequence of propositions 2.2 and 3.1.

**Proposition 5,2.**
Let M be a digital n- manifold, n>0, {x,y} be a simple pair lying in M, and N=(M∪z)-{x,y} be the space obtained by the contraction of {x,y}. Then the rim O(z)=U(x)∪U(y)-{x,y} is a digital (n-1)-sphere, and N=(M∪z)-{x,y} is a digital n- manifold homotopy equivalent to M.
The proof of this proposition is similar to the proof of proposition 3,1 and is omitted.

**Proposition 5.3.**
Let M be a digital n-manifold, and S be a digital (n-1)-sphere lying in M. M is a digital n-sphere if and only if S is a separating space of M=A∪S∪B, and A∪S and S∪B are digital n-disks.

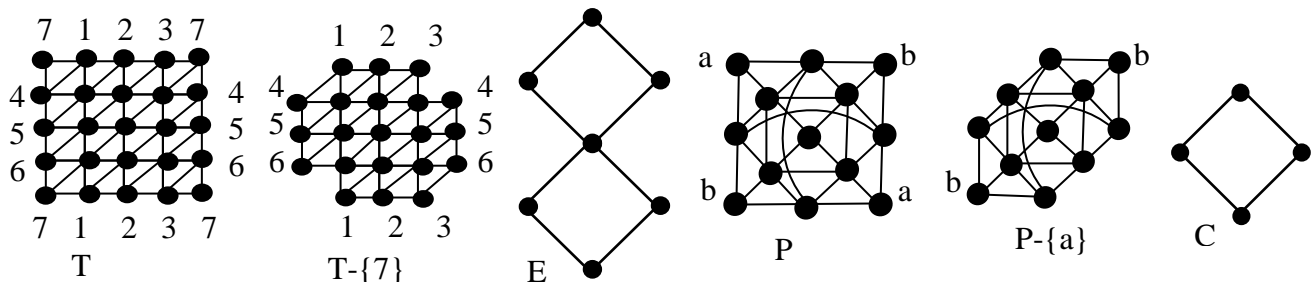

Figure 8. A digital 2-dimensional torus T and a digital 2-dimensional projective plane P. T-{7} and P-{a} are 2-manifolds with boundary O({7}) and O({a}) respectively. By sequential deleting simple points and edges, T-{7} can be converted to E, and P-{a} can be converted to C.

**Proof.**
Suppose that M is a digital n-sphere. Then S a separating space of M=A∪S∪B, and A∪S and S∪B are digital n-disks by proposition 4.1.
For the converse, suppose that S is a separating space of M=A∪S∪B, and A∪S and S∪B are digital n-disks, Int(A∪S)=A, Int(B∪S)=B. Suppose that |A|>1 and |B|>1. According to proposition 3.3, A can be transformed to a point x, and B can be transformed to a point y by sequential contractions of simple pairs. Since the obtained space N=$S^0$(x,y)⊕S is a digital n-sphere, then M is a digital n-sphere. The proof is complete. □

**Proposition 5.4.**
Let M be a digital 3-manifold. If for any point v∈M, |O(v)|≤7 then M is a digital 3-sphere with |M|≤10.
**Proof.**
It can be checked directly that if for any point v∈M, |O(v)|=6, then $M=S^3_{min}= S^0_1\ldots\oplus S^0_4$; if for some points v∈M, |O(v)|=6, and for some point u∈M, |O(u)|=7, then M is a digital 3-sphere $S^0_1\oplus S^0_2\oplus S^1(5)$; if for any point v∈M, |O(v)|=7, then $M= S^1(5)\oplus S^1(5)$, where $S^1(5)$ is a digital 1-sphere consisting of five points. □

## 6. Some properties of simply connected digital spaces and manifolds

Digital simple closed curves have been studied in a number of papers. In particular, it was shown in [1, 13] that a digital simple closed curve of more than four points is not contractible.
In classical topology, simple connectedness is a basic notion in the Poincare conjecture. According to J. Milnor [15], the Poincaré conjecture can be formulated as follows: If a smooth compact 3-dimensional manifold M has the property that every simple closed curve within the manifold can be deformed continuously to a point, does it follow that M is homeomorphic to the three-sphere S ?
A few years ago several groups presented papers that claimed to complete the proof of the Poincaré conjecture [see 2, 12, 16]. The results of these papers were based upon earlier papers by G. Perelman [17-19].
In topology, a topological space X is called simply connected if it is path-connected and for any continuous map $f : S^1 \to X$ there exists a continuous map $F : D^2 \to X$ such that $F(D^2)$ restricted to $S^1$ is $f(S^1)$. In a digital space, a digital simple closed curve C corresponds to a continuous simple closed curve and a digital 2-disk, which is a contractible space, corresponds to $F(D^2)$.
Taking into account this analogy, we give the following definitions.

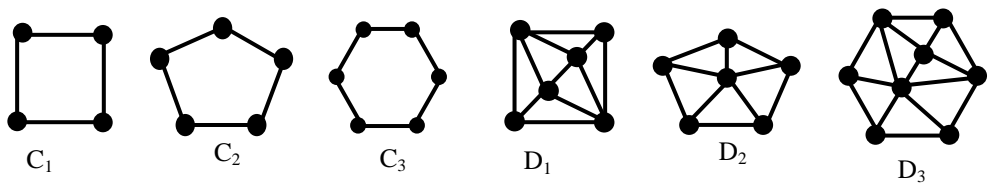

Figure 9. Simple closed curves $C_1$, $C_2$ and $C_3$. Digital 2-disks $D_1$, $D_2$ and $D_3$ with boundaries $C_1$, $C_2$ and $C_3$ respectively.

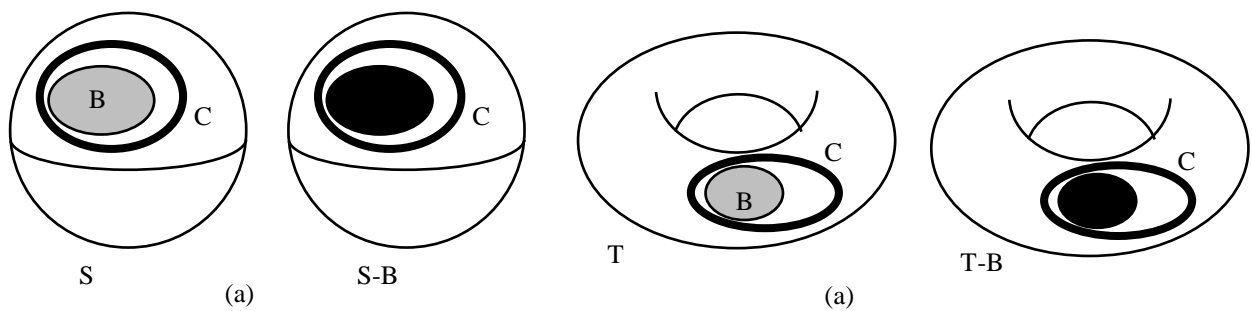

Figure 10. (a) S is a sphere, B is a topological disk lying in S, C is a simple closed curve lying in S-B, S-B is a simply connected space. (b) T is a torus, B is a topological disk lying in T, C is a simple closed curve lying in T-B. T-B is not a simply connected space.

**Definition 6.1.**
A simple closed curve is a digital 1-sphere.

**Definition 6.2.**
A digital space M is called simply connected if for any simple closed curve C contained in M, there is a digital 2-disk D=∂D∪IntD contained in M with the boundary C=∂D.

Remind that a digital 2-disk is a contractible space. Fig. 9 shows digital simple closed curves $C_1$, $C_2$ and $C_3$ and digital 2-disks $D_1$, $D_2$ and $D_3$ with boundaries $C_1$, $C_2$ and $C_3$ respectively. Digital n-disks and n-spheres, n>1, are simply connected (see fig. 3, 5). It is easy to check directly that a digital projective plane P, and a digital torus T depicted in fig. 8 are not simply connected.

**Definition 6.3.** A connected digital space M is called locally simply connected if for any contractible subspace B of M, the space M-B is a simply connected space.

If M is a locally simply connected digital space, then it is simply connected. Digital 2-disks depicted in fig. 5 are simply connected but not locally simply connected because for any point v belonging to IntD, D-v is not simply connected. All digital n-spheres, n>1, shown in fig. 3, 5 and 7 are locally simply connected spaces. It is not hard to check that a digital projective plane P and a digital torus T shown in fig. 8 are not a locally simply connected. In continuous case, the notion of local simple connectedness is quite clear. For illustration, consider closed surfaces depicted in fig. 10. Let S be a sphere, B be a topological disk lying in S, and C be a simple closed curve lying in S-B. Then S-B is simply connected, space. If T is a torus, B is a topological disk lying in T, and C is a simple closed curve lying in T-B, then T-B is not a simply connected, space, i.e., C can not be deformed continuously to a point within T-B. We may say that S is a locally simply connected surface, but T is not a locally simply connected surface.

**Theorem 6.1.**
Let M be a digital n-sphere. Then M is a locally simply connected space.
**Proof.**
The proof is by induction on the number $|M|=k$ of points of M. For $k=2n+2$, M is a minimal digital n-sphere $M=S^0_1(x_1,y_1)\oplus \ldots S^0_{n+1}(x_{n+1},y_{n+1})$ (see fig. 11). A simple closed curve $C\subseteq M$ contains exactly four points by the construction of M. Suppose that $C=S^0_1(x_1,y_1)\oplus S^0_2(x_2,y_2)$. By construction of M, a contractible space $B\subseteq M$, $C\cap B$, is a set of points without at least a point. Suppose that $x_3\notin B$. Then $x_3\oplus C$ is a digital 2-

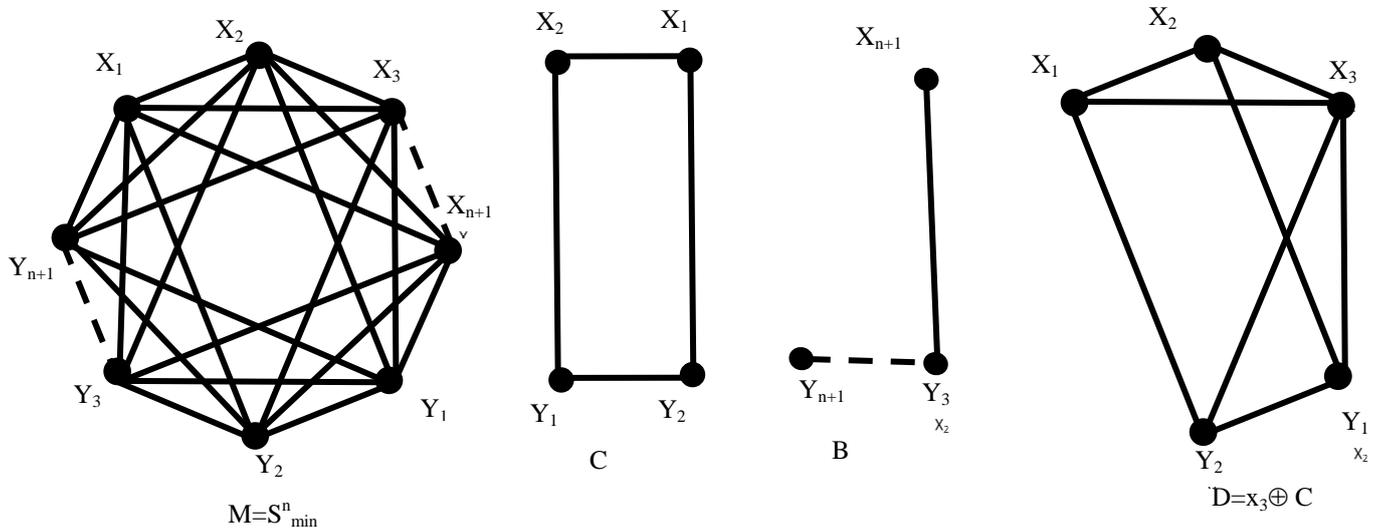

Figure 11. $M=S^0_1(x_1,y_1)\oplus \ldots S^0_{n+1}(x_{n+1},y_{n+1})$ is a minimal digital n-sphere, $C=S^0_1(x_1,y_1)\oplus S^0_2(x_2,y_2)$ is a simple closed curve, $B=\{x_4,x_5,\ldots x_{n+1}, y_1,\ldots y_{n+1}\}$ is a contractible space, $D=x_3\oplus C$ is a digital 2-disk with the boundary C.

disk such that $\partial D=C$.
Assume now that the assertion is valid whenever $2n+2<k<s$. Let $k=s$, B be a contractible subspace of M, and C be a simple closed curve lying in M-B. Then H=M-B is a contractible space and $C\subseteq H$ according to proposition 3.4. Since $|M|>2n+2$ then M contains a simple pair $\{x,y\}$. Let $N=FM=(M\cup z)-\{x,y\}$, $C_1=FC$ and $H_1=FH$ are digital spaces obtained by contraction of $\{x,y\}$. N, $C_1$ and $H_1$ are a digital n-sphere, a simple closed curve and a contractible space according to propositions 3.5 and 2.3. Then there is a digital 2-disk $D_1\subseteq H_1$ such that $\partial D_1=C_1$ by the induction hypothesis. Let $R=F^{-1}$ be the inverse of F. Then $D=RD_1$ is a digital 2-disk, $C=RC_1$ is a simple closed curve, $H=RH_1$, $D\subseteq H$ and $\partial D=C$. Therefore, M is a locally simply connected space. The proof is complete. ☐

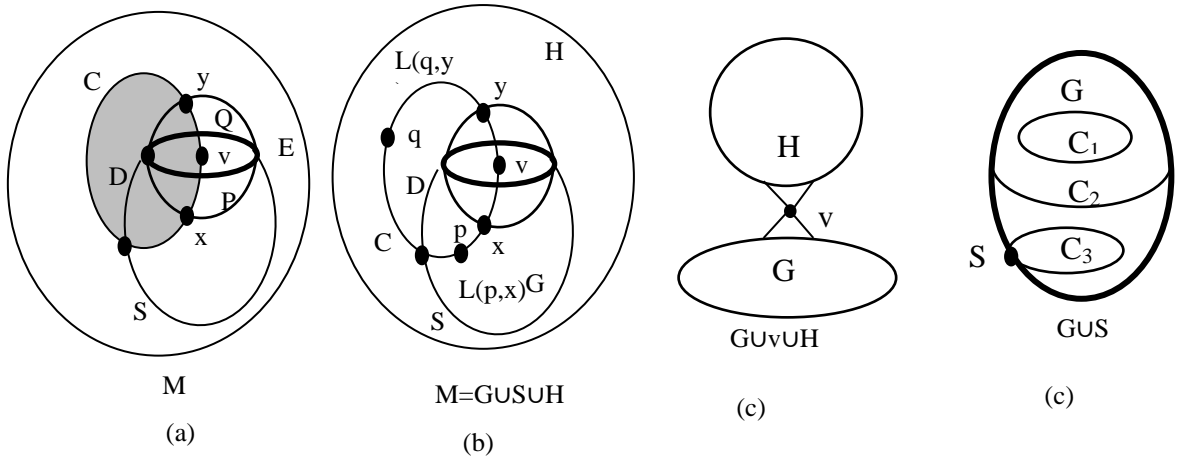

Figure 12. (a) M is a locally simply connected n-manifold, S is a digital (n-1)-sphere, point $v\in S$, $O(v)=G\cup E\cup H$, C is a simple closed curve, D is a digital 2-disk, $C=\partial D$. (b) C is a simple closed curve containing points {q,y,v,x,p}. (c) $G\cup v\cup H$ is a simply connected space. (d) $G\cup S$ is a simply connected space.

**Theorem 6.2.**
Let M be a digital n-manifold, n>0, {x,y} be a simple pair lying in M, and $FM=N=(M\cup z)-\{x,y\}$ be the space obtained by the contraction of {x,y}. M is locally simply connected iff N is locally simply connected.
**Proof.**
(1) Let M be locally simply connected, $B\subseteq N$ be a contractible space, and C be a simple closed curve lying in N-B. Then $C_1=F^{-1}C$ is a simple closed curve lying in M, and $B_1=F^{-1}B$ is a contractible space lying in M according to propositions 2.3 and 3.5. By construction of M and F, $C_1\subseteq M-B_1$. Therefore, there is a digital 2-disk $D_1\subseteq M-B_1$ such that $C_1=\partial D_1$. Hence, $CD_1=D\subseteq M-B$, and $\partial D=C$. Thus, N is a locally simply connected digital n-manifold.
(2) For the converse, suppose that N is locally simply connected, $B\subseteq M$ is a contractible space, and C is a simple closed curve lying in M-B. The proof that M is locally simply connected is similar to the proof above, and so is omitted. This completes the proof. ☐

**Theorem 6.3.**
Let M be a digital n-manifold and S be a digital (n-1)-sphere lying in M. If M is locally simply connected then S is a separating subspace of $M=A\cup S\cup B$, and $A\cup S$ and $S\cup B$ are digital n-manifolds with spherical boundary S.
**Proof.**
If S is the rim O(v) of some point v, then the assertion is plainly true. Let S is different from the rim of any point of M. Pick a point v belonging to S. S-v is a digital n-disk, i.e., a contractible space according to proposition 3.3. The rim O(v) is a digital (n-1)-sphere. Since $E=O(v)\cap S$ is a digital (n-2)-sphere, then E is a separating space of O(v) according to proposition 6.1, i.e., $O(v)=Q\cup E\cup P$ is a separation of O(v) (see fig. 12(a)).
Let points $x\in P$, $y\in Q$ and C be a simple closed curve containing points x, v and y.
Assume that $(S-v)\cap C=\emptyset$. Since M is locally simply connected, then there is a digital 2-disk D (dark area in fig 12 (a)) contained in M-(S-v) with the boundary $\partial D=C$, i.e., $D\cap(S-v)=\emptyset$. Since points x, y and v belong to D then $D\cap E\neq\emptyset$ by construction of O(v). Therefore, $D\cap(S-v)\neq\emptyset$. This is a contradiction. Thus, the assumption is false and $(S-v)\cap C\neq\emptyset$.
Suppose that for some point p there is a path L(p,x) such that $L(p,x)\cap S=\emptyset$, and G is the set of all such points (see fig. 12(b)). Suppose that for some point q there is a path L(q,y) such that $L(q,y)\cap S=\emptyset$, and H is the set of all such points. Consider a simple closed curve C containing points {p,x,v,y,q}. Since C intersects S-v, then points p and q are not adjacent. Hence, $M=G\cup S\cup H$ is a separation of M by S. $G\cup S$ and $S\cup H$ are digital n-manifolds with spherical boundary S by construction of M. The proof is complete. ☐

Fig.12 (a)-(b) illustrates theorem 6.3.

**Theorem 6.4.**
Let M be a locally simply connected digital n-manifold, S be a digital (n-1)-sphere lying in M, and M=G∪S∪H be a separation of M by S. Then subspaces G∪S and S∪H are both simply connected digital n-manifolds with spherical boundary S.
**Proof.**
Pick a point v belonging to S. According to theorem 3.3, B=S-v is a digital n-disk and a contractible space. Then M-B is a simply connected space. M-B can be represented as the union G∪v∪H, where v is a separating space of M-B. By construction of G∪v∪H, spaces G, H, G∪v and v∪H are all simply connected (see fig. 12 (c )).
Consider the union G∪S (fig. 12 (d )).
If a simple closed curve $C_1$ lies in G, then there is a digital 2-disk D lying in G∪v with the boundary $\partial D=C_1$.
If a simple closed curve $C_2$ lies in S, then there is a digital 2-disk D lying in S⊆G∪S with the boundary $\partial D=C_2$ according to theorem 6.1.
Suppose that C lies in G∪S and S∩C≠∅, G∩C≠∅, and D is a digital 2-disk with the boundary $\partial D=C$. Then there is a point v∈IntD∩G. Since IntD is a connected space, and S is a separating space of M then any point belonging to IntD must belong to G∪S. Thus, G∪S is a simply connected space. For the same reason as above, S∪H is a simply connected space. The proof is complete. □

**7. Digital locally simply connected 2- and 3-manifolds**

**Remark 7.1.**
Notice that properties of digital n-manifolds and their connection with continuous n-manifolds were studied in [8]. In particular, it was shown that if G and H are digital n-manifolds, and G is a subspace of H, then G=H.

**Theorem 7.1.**
If a digital 2-manifold M is locally simply connected, then M is a digital 2-sphere.
**Proof.**

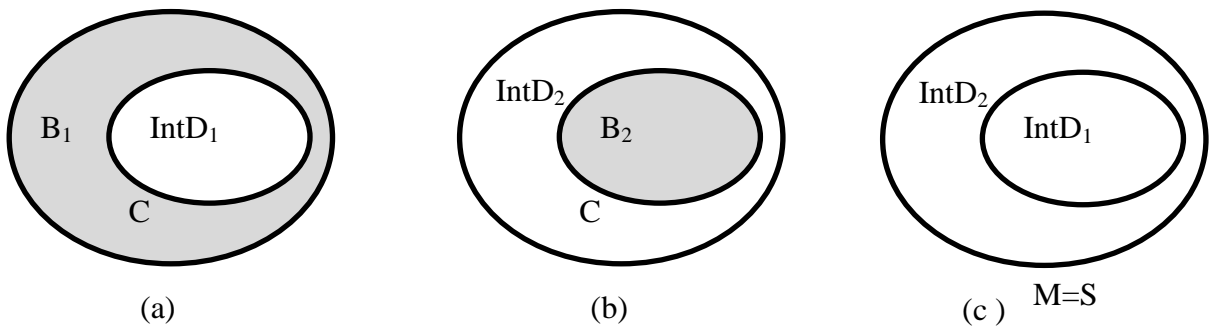

Figure 12. M a digital locally simply connected 2-manifold. M= $IntD_1 \cup C \cup IntD_2$ is a digital 2-sphere.

Let $B_1$ be a contractible subspace of M, and C be a simple closed curve contained in $M-B_1$ (fig 13(a). Then there is a digital 2-disk $D_1$ contained in $M-B_1$ such that $\partial D_1=C$. According to corollary 5.1, $IntD_1=B_2$ is a contractible space. Therefore, there is a digital 2-disk $D_2$ contained in $M-B_2$ such that $\partial D_2=C$ (fig. 13(b)). By construction, $IntD_1 \cup C \cup IntD_2$ is a digital 2-sphere S, and S⊆M. According to remark 7.1, S=M (fig. 13(c)). The proof is complete. □

**Theorem 7.2.**
If a digital 3-manifold M is locally simply connected, then M is a digital 3-sphere.
**Proof.**
Note first that if the rim of any point v belonging to M contains seven or less points, then M is a digital 3-

sphere according to proposition 5.4.

The proof is by induction on the number of points |M| in M. For |M|=8,9,10, M is a digital 3-sphere as it follows from proposition 5.4.

Assume now that the assertion is valid whenever |M|<k. According to theorem 5.3, that means that a simply connected digital 3-manifold N=∂N∪IntN with a spherical boundary, and the number of points |N|<k-1 is a digital 3-disk.

Let |M|=k. Consider a point v such that the number of points in the rim O(v) is more than seven, |O(v)|>7. Then O(v)=P∪C∪Q, where a simple closed curve C (see fig 14(a)) is a separating space of O(v), and |P|>1 and |Q|>1 according to proposition 4.2. According to proposition 2.1, B=v⊕(P∪Q) is a contractible space and C⊆M-B (see fig. 14(b)). Therefore, there exists a digital 2-disk D lying in M-B with the boundary C=∂D (see fig. 14(c)) according to definitions 6.2 and 6.3. Since S=D∪v is a digital 2-sphere, then S is a separating space in M=G∪S∪H, and G∪S and S∪H are simply connected digital 3-manifolds with the boundary S according to theorem 6.4. Since P⊆G and Q⊆H, then |G|>1, |H|>1. Since |M|=|G|+|H|+|S|=k, then |G|+|S|<k-1 and |H|+|S|<k-1. Hence, G∪S and S∪H are both digital 3-disks by the induction hypothesis. Thus, M=G∪S∪H is a digital 3-sphere by proposition 5.3. The proof is complete. □

A link between a continuous and a digital 3-manifold can be established by using the intersection graph of a cover of the 3-manifold (see [8]).

**Conclusions**
- A digital n-sphere M can be converted to the minimal n-sphere by sequential contractions of simple pairs.
- Let M be a digital n-manifold and v be a point of M. M is a digital n-sphere if and only if M-{v} is a digital n-disk.
- Let M be a digital n-sphere and S be a digital (n-1)-sphere lying in M. Then M=G∪S∪H is

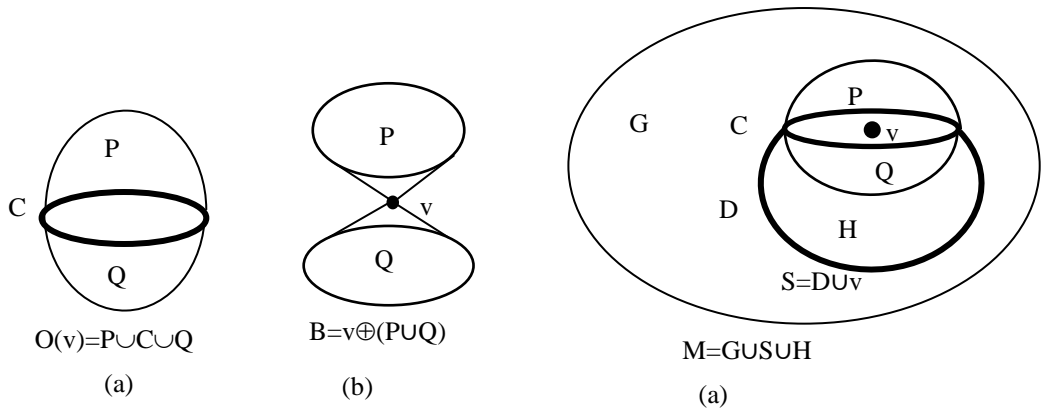

Figure 13. (a) O(v) is the rim of a point v, |P|>1, |Q|>1. (b) B=v⊕(P∪Q) is a contractible space contained in M, C⊆M-B. (c) S=D∪v is a separating space of M=G∪S∪H, G∪S and S∪H are digital 3-disks, M is a digital 3-sphere.

  the separation of M by S and G∪S and S∪H are digital n-disks.
- Let M be a digital n-manifold and S be a digital (n-1)-sphere lying in M. If M is locally simply connected then S is a separating subspace of M=G∪S∪H.
- If a digital 2-manifold M is locally simply connected, then M is a digital 2-sphere.
- If a digital 3-manifold M is locally simply connected, then M is a digital 3-sphere.

**Open problem**

Give a proof of theorem 7.2 for a digital locally simply connected n-manifold.

**Appendix 1: The Proofs**

**Proof of proposition 3.1.**

The proof is by induction on the number of points |M| of M. Let the cardinality $|M|=2n+3=|S^n_{min}|+1$. Then M contains a simple pair {x,y}, and $N=(M\cup z)-\{x,y\}$ is a minimal n-sphere by definition 3.2. Therefore, $U(x)\cup U(y)-\{x,y\}$ is a digital (n-1)-sphere because O(z) is a digital (n-1)-sphere lying in $N=S^n_{min}$, and $U(x)\cup U(y)-\{x,y\}=O(z)$ (see fig. 2, 5). Assume that the proposition is valid whenever |M|<2n+2+p, and let |M|=2n+2+p. Since $S^n_{min}=C_p\ldots C_1M$, then $M=R_1\ldots R_pS^n_{min}$, where $R_i=C_i^{-1}$, i=1,…p, according to definition 2.1. Hence $N=R_2\ldots R_pS^n_{min}$. The rim of any point belonging to N is a digital (n-1)-sphere by the induction hypothesis. Since z∈N then $U(x)\cup U(y)-\{x,y\}=O(z)$ is a digital (n-1)-sphere. □

**Proof of proposition 3.2.**
The proof is by induction on the number of points |M|=k of M. For k=2n+2, M is a minimal digital n-sphere, D is a minimal digital n-disk, i.e., a contractible space by construction (see fig. 3). Assume that the proposition is valid whenever k<s. Let k=s. Then M contains a simple pair{x,y}. With no loss of generality, suppose that v≠x, y. Evidently, $N=(M\cup z)-\{x,y\}$ is a digital n-sphere, |N|=s-1, v∈N. Therefore, $N-v=D_1$ is a contractible space by the induction hypothesis. Since $D_1=(D\cup z)-\{x,y\}$, then D and $D_1$ are homotopy equivalent, as it follows from propositions 2.2 and 2.3. Hence, D is a contractible space. The proof is complete. □

**Proof of proposition 3.3.**
The proof is by induction on the number of points |M|=k of M. For k=2n+2+1, the proposition is verified directly. (see fig. 2, 5).
Assume that the proposition is valid whenever k<s. Let k=s. M contains a simple pair{x,y}. Suppose that for a digital n-disk M-v=D=∂D∪IntD, |IntD|>1.
(1) Let v=x, y∈O(v), $N=(M\cup z)-\{x,y\}$, and N-z=E=∂E∪IntE.
(a) Suppose that |IntE|>1. Then IntE contains a simple pair by the induction hypothesis. Since IntE⊆IntD, then |IntD|>1.
(b) Suppose that |IntE|=1, i.e., IntE=u. Then for any point p∈U(y)-U(x), {u,p} is a simple pair by construction. Evidently, {u,p} lies in IntD.
(2) Suppose that {x,y}⊆O(v), $N=(M\cup z)-\{x,y\}$, and N-v=E=∂E∪IntE. Then |IntD|=|IntE|>1. IntE contains a simple pair {a,b} by the induction hypothesis. By consruction, {a,b} is a simple pair contained in IntD.
(3) Suppose that {x,y}⊆M-v. Then IntD contains {x,y}. The proof is complete. □

**Proof of proposition 3.5.**
(1) If {x,y}⊆G then {x,y} is a simple pair of G according to proposition 3.2. $G_1$=FG is a digital (n-k)-sphere according to proposition 3.1, and $G_1\subseteq N$. If {x,y}⊄G then $G_1$=G is a digital (n-k)-sphere lying in N by construction..
(2) Consider a digital n-sphere M=v∪D where the rim O(v)=∂D. Then {x,y} is simple pair of M by construction. Therefore, $N=FM=(M\cup z)-\{x,y\}$ is a digital n-sphere by proposition 3.1. By construction of N, $N-v=E=(D\cup z)-\{x,y\}$ is a digital n-disk. The proof is complete. □